\documentclass[twocolumn,secnumarabic,amssymb, nobibnotes, aps, prd]{revtex4}
\usepackage{graphicx}

\begin{document}
\title{Experimental investigations of the change with magnetic flux of quantum number in superconducting ring}
\author{A.V. Burlakov, V.L. Gurtovoi, A.I. Ilin, A.V.  Nikulov, and V.A. Tulin}
\affiliation{Institute of Microelectronics Technology and High Purity Materials, Russian Academy of Sciences, 142432 Chernogolovka, Moscow District, RUSSIA.} 
\begin{abstract} The magnetic dependencies of the critical current of aluminum ring with asymmetric link-up of current leads have been measured in order to clear up the essence of the paradoxical absence of the jump of the critical current at the quantum number change revealed before at the measurements of asymmetric superconducting ring. The measurements have shown that the experimental and theoretical dependencies agree in the region of magnetic field corresponding to integer numbers of the flux quantum and disagree at the half of the flux quantum. The jump is not observed as well as in the asymmetric rings. 
 \end{abstract}

\maketitle

\narrowtext

\section*{Introduction}

The discreteness of the spectrum of the permitted velocity $v = p/m $ states because of the Bohr's quantization $pr  = n\hbar $ goes down with the radius $r$ and mass $m$ increase, in agreement with the Bohr's correspondence principle. Therefore the quantum phenomena, such as the persistent current of electrons, can be observed in nano-rings with $r > 300 \ nm$ only at low temperature \cite{Nikulov01,Nikulov02} even at microscopic mass $m = 9 \ 10^{-31} \ kg$. Superconductivity is the exception thanks to the impossibility for Cooper pairs to change their quantum state $n$ individually \cite{Nikulov03}. This exception gives us an opportunity for more detailed research of phenomena connected with the Bohr's quantization: a ring, in contrast to atom, can be made in different forms. 

\section {Periodical dependencies in magnetic field}
We can observe the change of the quantum number $n$ in magnetic field $B$ thanks to a "huge" radius $r > 500 \ nm$ of  superconducting ring in comparison with atom orbit radius $r_{B} \approx 0.05 \ nm $. This change is observed because of the Aharonov-Bohm effect \cite{Nikulov01,Nikulov04}, i.e. the influence of the magnetic vector potential $A$ of the canonical momentum $p = mv + qA$ of a particle with charge $q$. The momentum of a free particle is described with the gradient $p = \hbar \bigtriangledown \varphi $ of  the phase $\varphi $ of the wave function $\Psi = |\Psi |\exp{i\varphi }$, according to the quantum formalism. Because of the requirement that the complex wave function, describing a particle state, must be single-valued $\Psi  = |\Psi |\exp i\varphi = |\Psi |\exp i(\varphi + n2\pi)$ at any point, the phase $\varphi $ must change by integral $n$ multiples of $2\pi $ following a complete turn along the path of integration, yielding $\oint_{l}dl \bigtriangledown \varphi = \oint_{l}dl p/\hbar = \oint_{l}dl (mv + qA)/ \hbar = m\oint_{l}dl v/ \hbar + 2\pi \Phi/\Phi_{0}= n2\pi $. The integer number $n$ of the permitted state with minimum energy $\propto v^{2} \propto (n - \Phi/\Phi_{0})^{2}$ should change with magnetic flux $ \Phi = BS \approx B\pi r^{2}$ at $ \Phi = (n+0.5)\Phi_{0}$. The flux quantum $\Phi _{0} = 2\pi \hbar /q $ equals $\approx 4140 \ T \ nm^{2}$ for electron $q = e$ and $\approx 2070 \ T \ nm^{2}$ for Cooper pair $q = 2e$. Consequently the $n$ change can be observed in magnetic field $B > \Phi_{0}/ \pi r^{2} \approx 0.0026 \ T $ in a superconducting ring with $r \approx 500 \ nm$ and in the inaccessibly high magnetic fields $\Phi_{0}/ \pi r_{B}^{2} \approx 530000 \ T$ for atom orbit with $r_{B} \approx 0.05 \ nm $. 

The magnetic periodicity in different parameters, such as: resistance \cite{Nikulov04,Nikulov05,Nikulov06,Nikulov07}, magnetic susceptibility \cite{Nikulov08}, critical current \cite{Nikulov09,Nikulov10} and dc voltage \cite{Nikulov06,Nikulov07,Nikulov09} demonstrate unambiguously the $n$ change with the magnetic flux $ \Phi = BS $ inside ring: the resistance oscillations have minimum value at $ \Phi = n\Phi_{0}$ and maximum value at $ \Phi = (n+0.5)\Phi_{0}$, in accordance with the theoretical relation $\Delta R \propto v^{2} \propto (n - \Phi/\Phi_{0})^{2} $; magnetic susceptibility and dc voltage change the sign at $ \Phi = n\Phi_{0}$ and $ \Phi = (n+0.5)\Phi_{0}$, in accordance with the theoretical relation $M \propto V_{p} \propto \overline{v} \propto \overline {n - \Phi/\Phi_{0}}$; the critical current of symmetric ring have minimum value at $ \Phi = (n+0.5)\Phi_{0}$ and maximum value at $ \Phi = (n+0.5)\Phi_{0}$, in accordance with the relation $I_{c} = I_{c0} - 2I_{p}$, where $I_{p} = s2en_{s}v =    I_{p,A}2(n - \Phi /\Phi _{0})$ is the persistent current circulating in the ring clockwise or anticlockwise \cite{Nikulov09}. No jump, which could be connected with the jump of the permitted velocity $v \propto n - \Phi/\Phi_{0}$ and the persistent current, because of the $n$ change at $ \Phi = (n+0.5)\Phi_{0}$ is observed at these measurements in accordance with the theory: the squared velocity $v^{2} \propto (n - \Phi/\Phi_{0})^{2}$ and the absolute value of the persistent current $|I_{p}| =    I_{p,A}2|n - \Phi /\Phi _{0}| $ should not change with $n$ change from $n$ to $n+1$ at $ \Phi = (n+0.5)\Phi_{0}$; the average velocity at $\overline{v} \propto \overline {n - \Phi/\Phi_{0}} = (-0.5 + 0.5)/2 = 0$ at $ \Phi = (n+0.5)\Phi_{0}$. 

\begin{figure}[]
\includegraphics{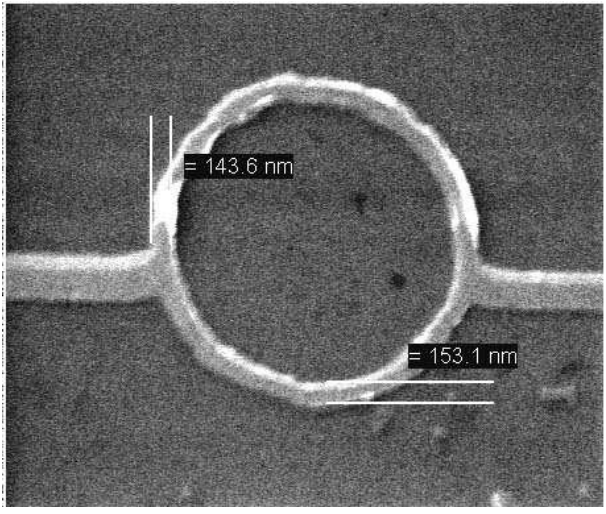}
\caption{\label{fig:epsart} The aluminum ring $r \approx 1 \mu m$ with asymmetric link-up of current leads, $l_{sh} \approx 0.7\pi r$, $l_{long} \approx 1.3\pi r$. }
\end{figure}

\begin{figure}[b]
\includegraphics{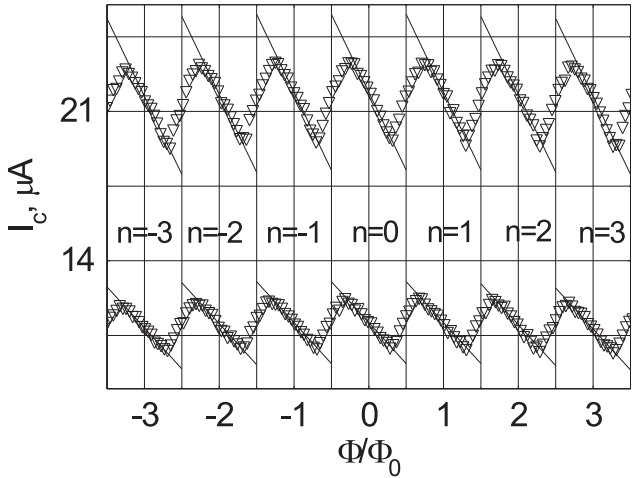}
\caption{\label{fig:epsart} The magnetic dependencies of the critical current $I_{c+}(\Phi /\Phi _{0})$ of aluminum ring with asymmetric link-up of current leads measured at the temperature $T \approx 0.900T_{c}$; $T \approx 0.933T_{c}$; $T_{c} \approx 1.52 \ K$. }
\end{figure}

\section {The jump absence at measurements of the critical current with different section of ring halves}
The jump could be expected at measurement of the oscillations of the critical current of a superconducting ring with different section of ring halves $ s_{w} > s_{n}$. The critical current measured in the one $I_{c+}$ or other $I_{c-}$ direction should be equal \cite{Nikulov09}
$$I_{c+}, I_{c-} = I_{c0} - I_{p,A}2|n - \frac{\Phi }{\Phi _{0}}|(1 + \frac{s_{w}}{s_{n}}) \eqno{(1a)}$$
when the external current $I_{ext}$ measuring $I_{c+}$, $I_{c-}$ is added with the persistent current $I_{p}$ in the narrow ring half $ s_{n}$and
$$I_{c+}, I_{c-} = I_{c0} - I_{p,A}2|n - \frac{\Phi }{\Phi _{0}}|(1 + \frac{s_{n}}{s_{w}}) \eqno{(1b)}$$
when the currents are added in the wide ring half $ s_{w}$. The jump $\Delta I_{c} = I_{p,A}( s_{w}/s_{n} - s_{n}/s_{w})$ should be observed because of the change of the persistent current direction with the $n$ change at $ \Phi = (n+0.5)\Phi_{0}$. But the measurements \cite{Nikulov09,Nikulov10,Nikulov11} have revealed the absence of the expected jump in spite of the enough large amplitude $ I_{p,A}$ of the persistent current and the relation $s_{w}/s_{n} \approx  2$. According to the universally recognised quantum formalism the anisotropy $I_{c,an}(\Phi /\Phi _{0}) = I_{c+}(\Phi /\Phi _{0}) - I_{c-}(\Phi /\Phi _{0})$ of the critical current measured in the opposite directions should appear in the asymmetric ring $s_{w}/s_{n} > 1$ because of the change of the functions $I_{c+}(\Phi /\Phi _{0})$, $I_{c-}(\Phi /\Phi _{0})$ describing its magnetic dependencies, see Fig.19 \cite{Nikulov09} and  Fig.3 \cite{Nikulov11}. But it was found \cite{Nikulov10} that the anisotropy $I_{c,an}(\Phi /\Phi _{0}) = I_{c}(\Phi /\Phi _{0}+ \Delta \phi) - I_{c}(\Phi /\Phi _{0} - \Delta \phi)$ appears because of the changes in the arguments $ I_{c+}(\Phi /\Phi _{0}) \approx I_{c}(\Phi /\Phi _{0}+ \Delta \phi)$, $ I_{c-}(\Phi /\Phi _{0}) \approx I_{c}(\Phi /\Phi _{0}- \Delta \phi)$ of the functions rather than the functions themselves. The shift of the periodic dependencies of the critical current on the quarter of the flux quantum $\Delta \phi \approx 1/4$ \cite{Nikulov10} is very paradoxical phenomenon which can not be explained now as well as the absence of the $I_{c+}$,  $I_{c-}$ jump with the $n$ change. 

\section {Measurements of the critical current of superconducting rings with asymmetric link-up of current leads}
In order to clear up the essence of the second puzzle we have measured the magnetic dependencies of the critical current of aluminum ring with asymmetric link-up of current leads, Fig.1. The critical current should determined with the transition into the resistive state only of the shot segment $l_{sh} \approx 0.35 \times 2\pi r$ at the relation of the persistent current amplitude and the critical current $I_{p,A}/I_{c0} < 0.25$ measured on the ring, Fig.2, with the segment relation $l_{long}/l_{sh} \approx 0.65/0.35  \approx 1.8$. The critical current dependencies should describe with the relations 
$$I_{c+} = I_{c0} - I_{p,A}(n - \frac{\Phi }{\Phi _{0}}) \eqno{(2a)}$$
$$I_{c-} = I_{c0} + I_{p,A}(n - \frac{\Phi }{\Phi _{0}}) \eqno{(2b)}$$
implying the jump $\Delta I_{c} = I_{p,A}$ at $ \Phi = (n+0.5)\Phi_{0}$, Fig.2. The measured dependencies $I_{c+}(\Phi /\Phi _{0})$,  $I_{c-}(\Phi /\Phi _{0})$ correspond to the expected one (2) in the region near $ \Phi = n\Phi_{0}$, Fig.2. Nevertheless the jump, which must be because of the quantum number $n$ change according to (2), is not observed, Fig.2. The measured values of the critical current $I_{c+}(\Phi /\Phi _{0})$,  $I_{c-}(\Phi /\Phi _{0})$ vary smoothly with magnetic flux value near $ \Phi = (n+0.5)\Phi_{0}$, Fig.2, in defiance of the quantum formalism predicting the $I_{c}$ jump (2). The aspiration of Nature to avoid the jump surprises. It is puzzle that the experimental and theoretical dependencies agree near $ \Phi \approx  n\Phi_{0}$ and disagree near $ \Phi \approx  (n+0.5)\Phi_{0}$.

\end{document}